\documentclass[PAPER, cernpreprint, atlasdraft=false, texlive=2023, USenglish, texmf, orcidlogo]{atlasdoc}
 
\usepackage[subcaption=true]{atlaspackage}
\usepackage{atlasbiblatex}
 
\usepackage{atlasphysics}

\graphicspath{{logos/}{figures/}}
 
\usepackage{multirow}
\usepackage{multicol}
\usepackage{nicematrix}

\AtlasTitle{Observation of $W\gamma\gamma$ triboson production in proton-proton collisions at $\sqrt{s} = 13$~TeV with the ATLAS detector}

\AtlasAbstract{ 
This letter reports the observation of $W(\ell\nu)\gamma\gamma$ production in proton-proton collisions. This measurement uses the full Run~2 sample of events recorded at a \com energy of $\sqrt{s} = 13$~TeV by the ATLAS detector at the LHC, corresponding to an integrated luminosity of 140~fb$^{-1}$.   Events with a leptonically-decaying $W$ boson and at least two photons are considered. The background-only hypothesis is rejected with an observed and expected significance of $5.6$ standard deviations. The inclusive fiducial production cross section of $W(e\nu)\gamma\gamma$ and $W(\mu\nu)\gamma\gamma$ events is measured to be $\sigma_{\mathrm{fid}} = 13.8 \pm 1.1 (\mathrm{stat}) \substack{+2.1 \\ -2.0} (\mathrm{syst}) \pm 0.1 (\mathrm{lumi})$~fb, in agreement with the Standard Model prediction.
}

\AtlasRefCode{STDM-2018-33}
 
\PreprintIdNumber{CERN-EP-2023-037}

\AtlasJournal{Phys.\ Lett.\ B.}
\AtlasJournalRef{\PLB 848 (2024) 138400}
\AtlasDOI{https://doi.org/10.1016/j.physletb.2023.138400}

\hypersetup{pdftitle={ATLAS document},pdfauthor={The ATLAS Collaboration}}
 
\newcommand*{\etcone}{\ensuremath{\ET^{\text{cone}}}\xspace}
\newcommand*{\etconegen}{\ensuremath{\ET^{\text{cone, gen.}}}\xspace}
\newcommand*{\etisoy}{\ensuremath{\ET^{\text{iso}, \gamma}}\xspace}
\newcommand*{\etiso}{\ensuremath{\ET^{\text{iso}}}\xspace}
\newcommand*{\ptiso}{\ensuremath{\pT^{\text{iso}}}\xspace}
\newcommand*{\ptisovare}{\ensuremath{p_{\text{T,}e}^{\text{iso}}}\xspace}
\newcommand*{\ptisovarmu}{\ensuremath{p_{\text{T,}\mu}^{\text{iso}}}\xspace}

\newcommand*{\Loose}{\textit{Loose}\xspace}
\newcommand*{\Medium}{\textit{Medium}\xspace}
\newcommand*{\Tight}{\textit{Tight}\xspace}
\newcommand*{\com}{center-of-mass\xspace}
 
\newcommand{\AtlasCoordFootnote}{
ATLAS uses a right-handed coordinate system with its origin at the nominal interaction point (IP)
in the center of the detector and the \(z\)-axis along the beam pipe.
The \(x\)-axis points from the IP to the center of the LHC ring,
and the \(y\)-axis points upwards.
Cylindrical coordinates \((r,\phi)\) are used in the transverse plane,
\(\phi\) being the azimuthal angle around the \(z\)-axis.
The pseudorapidity is defined in terms of the polar angle \(\theta\) as \(\eta = -\ln \tan(\theta/2)\).
Angular distance is measured in units of \(\Delta R \equiv \sqrt{(\Delta\eta)^{2} + (\Delta\phi)^{2}}\).}
\begin{document}
 
\maketitle

\clearpage
 
\section{Introduction}
\label{sec:intro}
 
In the Standard Model (SM) of particle physics, interactions amongst
electroweak gauge bosons ($\gamma$, $W$, $Z$) are entirely determined by the non-Abelian
$\mathrm{SU(2)}\times\mathrm{U(1)}$ structure of the electroweak
sector.
In particular, in proton-proton collisions, the production of a $W$ boson
in association with two photons is sensitive to triple and
quartic gauge boson couplings that could be modified by the presence
of new physics phenomena~\cite{DEGRANDE201321,
Falkowski_2017, _boli_2004}.
The study of this process therefore provides
sensitivity to new physics that is complementary to direct searches as it can constrain new physics at energy scales that are beyond the reach of the LHC.
In addition, due to the small production cross section of the $W\gamma\gamma$ final state in proton-proton collisions, it is only now becoming accessible with the data collected during \RunTwo of the LHC. Therefore, it remains one of the least studied processes in the electroweak sector of the SM.
The production of a $W$ boson
in association with two photons is also an important background in a number of other
measurements, such as the production of the SM Higgs boson in association
with a $W$ boson, followed by a $H\rightarrow\gamma\gamma$ decay~
\cite{HIGG-2020-16}.
 
\begin{figure}[ht]
\centering
\begin{subfigure}[ht]{0.45\textwidth}
\centering
\includegraphics[width=\textwidth,keepaspectratio]{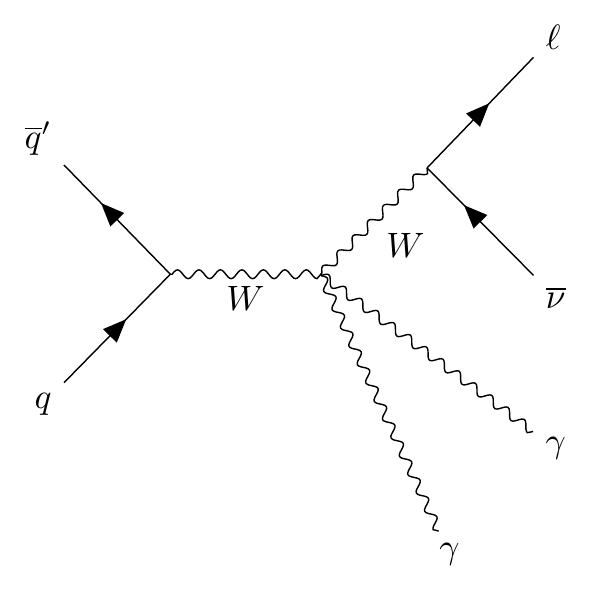}
\caption{$W\gamma\gamma$ production via a quartic gauge boson coupling.}
\label{feyn-qgc}
\end{subfigure}~~~~~~~~
\begin{subfigure}[ht]{0.45\textwidth}
\centering
\includegraphics[width=\textwidth,keepaspectratio]{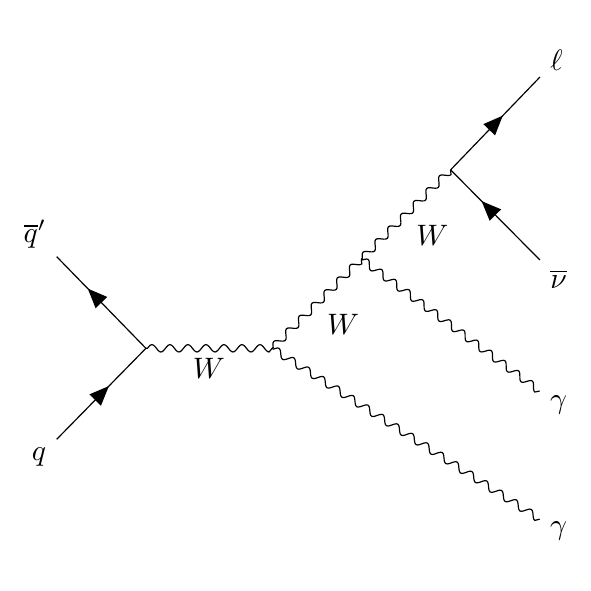}
\caption{$W\gamma\gamma$ production via two triple gauge boson couplings.}
\label{feyn-tgc}
\end{subfigure}\\
\begin{subfigure}[ht]{0.45\textwidth}
\centering
\includegraphics[width=\textwidth,keepaspectratio]{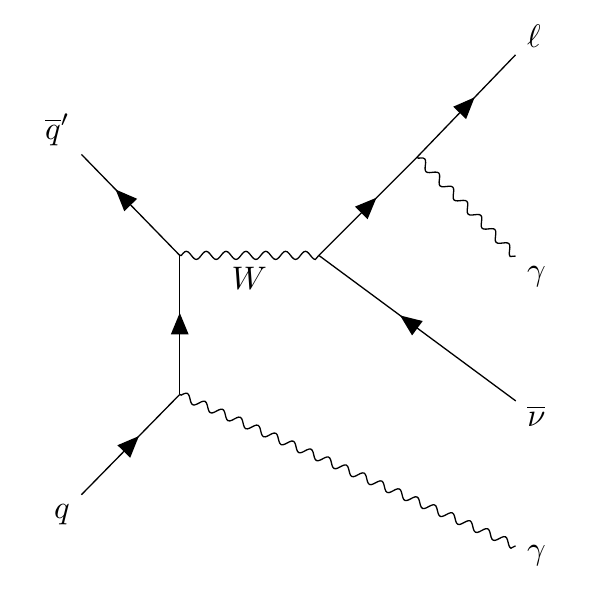}
\caption{$W\gamma\gamma$ production where the photons are produced via ISR and FSR.}
\label{feyn-isr-fsr}
\end{subfigure}~~~~~~~~
\begin{subfigure}[ht]{0.45\textwidth}
\centering
\includegraphics[width=\textwidth,keepaspectratio]{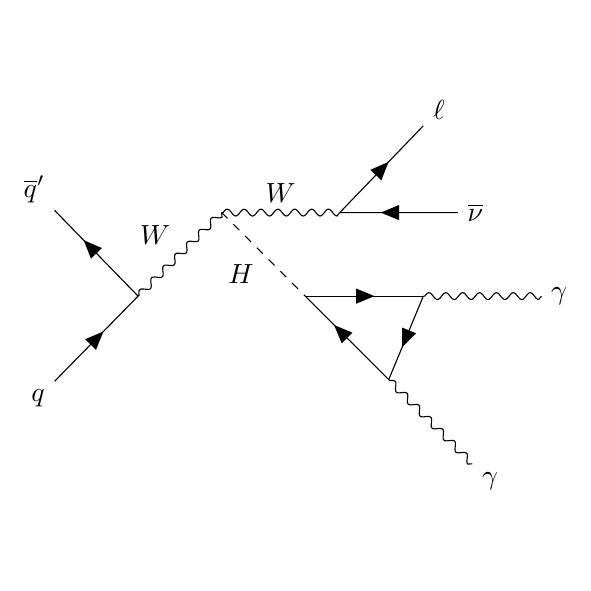}
\caption{$W\gamma\gamma$ production where the photons are produced via the decay of a SM Higgs boson.}
\label{feyn-higgs}
\end{subfigure}
\caption{Representative Feynman diagrams for the production of $W\gamma\gamma$.}
\label{feyn}
\end{figure}

The $W\gamma\gamma$ triboson production is studied here through final
states compatible with a leptonic decay of the $W$ boson. A representative selection
of leading-order (LO) Feynman diagrams, and a loop-induced SM Higgs boson Feynman diagram, of $pp\rightarrow
\ell\nu\gamma\gamma$ production are shown in Figure~\ref{feyn}.  These
Feynman diagrams illustrate four of the many possible production modes, and include processes where the photons are
produced via: \subref{feyn-qgc}~a $WW\gamma\gamma$ quartic gauge
coupling; \subref{feyn-tgc}~two $WW\gamma$ triple gauge couplings;
\subref{feyn-isr-fsr}~initial (ISR) and final (FSR) state radiation;
and \subref{feyn-higgs}~as the decay products of a Higgs boson. Production of $W\gamma\gamma$ via a SM Higgs boson is treated as background in this analysis to isolate the signal processes to those with only electroweak gauge boson interactions.
 
Although there are
contributions from processes with one or more FSR photons (see diagram~\subref{feyn-isr-fsr}), the process will nevertheless be referred to as
$W\gamma\gamma$ throughout this letter for simplicity.
 
The largest sources of background in this analysis consist of events in which at least one of the reconstructed objects in the final state is misidentified.  Data-driven techniques, described in Section~\ref{sec:background}, are used to estimate these sources of reducible background, which include photons from misidentified jets or neutral hadron decays, electrons misidentified as photons, leptons from misidentified jets or heavy-flavored hadron decays, and events in which one or both photons do not originate from the primary vertex. In addition, a small fraction of background events originates from multiboson ($WH(\gamma\gamma)$, $WW\gamma$, $Z\gamma\gamma$) and top-quark production ($tt\gamma$, $tW\gamma$, $tq\gamma$). Monte Carlo (MC) simulated samples, described in Section~\ref{sec:data}, are used to estimate the yield of these sources of irreducible background.
To maximize the analysis sensitivity, the uncertainty on background yield from $tt\gamma$ production is constrained from data in a control region (TopCR) that does not overlap with the signal region of interest.
 
Previous measurements of the $W\gamma\gamma$ process were performed at
the LHC using proton-proton collisions at a \com  energy of
$\sqrt{s}=8$~TeV with the ATLAS~\cite{STDM-2013-05} and
CMS~\cite{CMS-SMP-15-008} detectors, and at $\sqrt{s}=13$~TeV with the
CMS~\cite{CMS-SMP-19-013} detector, resulting in a maximum observed statistical
significance of $3.1\sigma$. This letter presents the
observation of the $W\gamma\gamma$ process
and a measurement of its fiducial cross section
in the $W\rightarrow e\nu$ and $W\rightarrow \mu\nu$
decay channels. In order to obtain a precise background estimate, the electron and muon channels are combined for both the observation and fiducial cross section measurement.
The $pp\rightarrow W\gamma\gamma$ signal strength $\mu$, defined as the ratio of the observed signal yield to the expected yield, is measured to assess the compatibility between data and SM prediction.
Results are obtained based on the analysis of
$\mathcal{L}=140~\mathrm{fb}^{-1}$ of proton-proton collision data collected with the ATLAS detector at $\sqrt{s}=13$~TeV, allowing for improvement over the previous ATLAS result due to both the increase in integrated luminosity and the increase in production cross section, in addition to improvements to the data-driven background estimates.
 
\section{ATLAS Detector}
The ATLAS experiment~\cite{PERF-2007-01} at the LHC is a multipurpose particle detector
with a cylindrical geometry, forward--backward symmetric, and a near \(4\pi\) coverage in
solid angle.\footnote{\AtlasCoordFootnote}
It consists of an inner tracking detector (ID) surrounded by a thin superconducting solenoid
providing a \SI{2}{\tesla} axial magnetic field, electromagnetic and hadron calorimeters, and a muon spectrometer (MS).
The inner tracking detector covers the pseudorapidity range \(|\eta| < 2.5\).
It consists of silicon pixel, silicon microstrip, and transition radiation tracking detectors.
Lead/liquid-argon (LAr) sampling calorimeters provide electromagnetic (EM) energy measurements
with high granularity.
A steel/scintillator-tile hadron calorimeter covers the central pseudorapidity range (\(|\eta| < 1.7\)).
The endcap and forward regions are instrumented with LAr calorimeters
for both the EM and hadronic energy measurements up to \(|\eta| = 4.9\).
The muon spectrometer surrounds the calorimeters and is based on
three large superconducting air-core toroidal magnets with eight coils each.
The magnetic field line integral of the toroidal magnets ranges between \num{2.0} and \SI{6.0}{\tesla\metre}
across most of the detector.
The muon spectrometer includes a system of precision tracking chambers and fast detectors for triggering.
A two-level trigger system is used to select events.
The first-level trigger is implemented in hardware and uses a subset of the detector information
to accept events at a rate below \SI{100}{\kHz}.
This is followed by a software-based trigger that
reduces the accepted event rate to \SI{1}{\kHz} on average
depending on the data-taking conditions.
An extensive software suite~\cite{ATL-SOFT-PUB-2021-001} is used in the reconstruction and analysis of real
and simulated data, in detector operations, and in the trigger and data acquisition systems of the experiment.

\section{Data and Simulation}
\label{sec:data}
 
The measurement presented in this letter is based on proton-proton collision data at a \com energy of $13$~TeV recorded by the ATLAS detector during \RunTwo of the LHC (2015-2018).  During this data-taking period, the number of interactions per proton bunch crossing (pileup) averaged between 13 and 38 interactions, depending on the year~\cite{ATLAS-CONF-2019-021}.  After applying ATLAS data quality requirements~\cite{DAPR-2018-01}, the dataset corresponds to an integrated luminosity of $\mathcal{L}=140$~fb$^{-1}$. The uncertainty in the combined integrated luminosity for 2015-2018 is 0.83\% \cite{DAPR-2021-01}, obtained using the LUCID-2 detector \cite{LUCID2} for the primary luminosity measurements, complemented by measurements using the ID and calorimeters.
 
Simulated samples are used to model the expected signal and irreducible background yields, while reducible backgrounds from misidentified objects are estimated using data-driven techniques described in Section~\ref{sec:background}. Some of the irreducible backgrounds, as listed in Section~\ref{sec:intro}, contribute to the analysis only when one lepton is not reconstructed or additional photons are present due to FSR.
 
Signal $e\nu\gamma\gamma$, $\mu\nu\gamma\gamma$, and $\tau\nu\gamma\gamma$ processes are generated with \textsc{Sherpa}~2.2.10~\cite{Bothmann:2019yzt} generator using next-to-leading-order (NLO) matrix elements (ME) with zero partons, and leading-order (LO) matrix elements for up to two partons calculated with the Comix~\cite{Gleisberg:2008fv} and OpenLoop~\cite{Cascioli:2011va,Buccioni:2019sur,Denner:2016kdg} libraries. They were matched with the \textsc{Sherpa} parton shower~\cite{Schumann:2007mg} using the MEPS@NLO prescription~\cite{Catani:2001cc,Hoeche:2009rj,Hoeche:2011fd,Hoeche:2012yf} using the set of tuned parameters developed by the \textsc{Sherpa} authors.
The \NNPDF[3.0nnlo] next-to-next-to-leading order (NNLO) parton distribution function (PDF) set from the NNPDF Collaboration~\cite{Ball:2014uwa} was used.  Approximate NLO electroweak corrections are included in these samples~\cite{Kallweit:2014xda} and result in a negligible effect in the $W\gamma\gamma$ phase space used in this measurement.
 
The $WH(\gamma\gamma)$ background is estimated from events generated with the \textsc{POWHEG-BOX~v2}~\cite{Alioli:2010xd} generator interfaced with \textsc{Pythia 8.212}~\cite{Sjostrand:2007gs, Mrenna:2016sih} using the AZNLO tune~\cite{STDM-2012-23} for parton showering modeling and NLO PDFs from NNPDF3.0. Background contributions from $WW\gamma$ (\textsc{Sherpa 2.2.11}), $Z\gamma\gamma$ (\textsc{Sherpa 2.2.10}), and $Z\gamma$ (\textsc{Sherpa 2.2.8}) processes are estimated from samples generated to NLO accuracy in perturbative QCD with up to 1 additional parton emission, and merged with samples to LO accuracy in perturbative QCD with 2 to 3 parton emissions; like the signal samples, these are generated with the NNLO PDF set from \NNPDF[3.0nnlo]. Double counting between $Z\gamma$ and $Z\gamma\gamma$ is removed at the event generation stage.
Contributions from $t\bar{t}\gamma$ events, $tW\gamma$ events where the photon is produced in the $tW$ decay chain, and $tq\gamma$ events are generated with \textsc{MadGraph5\_aMC@NLO 2.3.3}~\cite{Alwall:2014hca}. The \NNPDF[2.3lo]~\cite{Ball:2012cx} PDF sets and parton shower modeling from \textsc{Pythia 8.212} with the A14 tune~\cite{ATL-PHYS-PUB-2014-021} are used in the generation of these event samples. Contributions from $tW\gamma$ events where the photon is produced at matrix-element level are generated with \textsc{MadGraph5\_aMC@NLO 2.6.7}, \NNPDF[2.3lo] PDF sets, and parton shower modeling from \textsc{Pythia 8.244} with the A14 tune. In all simulated samples where the production of one photon is generated in the matrix element, a second prompt photon can be produced as FSR in the parton shower.
 
Both signal and background MC events are processed through the full ATLAS detector simulation~\cite{SOFT-2010-01} based on \textsc{GEANT4}~\cite{Agostinelli:2002hh}.
The effects of multiple interactions in the same and neighboring bunch
crossings are modeled by overlaying the simulated hard-scattering event with
inelastic proton--proton events generated with \PYTHIA[8.186]~\cite{Sjostrand:2007gs}
using the \NNPDF[2.3lo] set of PDFs~\cite{Ball:2012cx} and the
A3 set of tuned parameters~\cite{ATL-PHYS-PUB-2016-017}. Simulated events are weighted such that the pileup distribution reproduces the pileup distribution of the dataset used in this measurement.

\section{Event Selection}
 
The $W\gamma\gamma$ process is investigated using the leptonic decays of the $W$ boson. While events with a leptonic $\tau$ decay to an electron or a muon are considered as signal events, those with hadronic $\tau$ decays are considered as a background.
Candidate $W(\ell\nu)\gamma\gamma$ events therefore contain two isolated photons, an isolated electron or muon, and missing transverse momentum, with magnitude referred to as missing transverse energy (\MET), from the undetected neutrino(s) originating from the leptonic $W$ boson decays.
The following paragraphs describe the selection requirements used to define the signal region (SR) of the measurement.
 
Events used for this measurement are selected using a suite of triggers that require the presence of at least two photons with \pt > 10~GeV and at least one electron or muon with \pt > 20~GeV~\cite{TRIG-2018-01, TRIG-2018-05}. For the 2017-2018 data-taking period, the \pt thresholds used to select events with at least two photons and at least one electron were increased to 12~GeV (photons) and 24~GeV (electron). In addition to these triggers, single and di-lepton triggers with \pt thresholds between 14 and 26~GeV are used to select events for the data-driven background estimates. The overall efficiencies for these triggers to select simulated signal events in the signal region are 95\% in the electron channel and 82\% in the muon channel.
In all cases, trigger objects must be matched to reconstructed objects selected for analysis.
 
Events are required to have a primary vertex associated with at least two charged-particle tracks with $p_{\mathrm{T}}>0.5$~GeV in the proton-proton interaction region. If multiple vertices satisfy this criteria, the vertex with the highest $p_{\mathrm{T}}^2$ sum is selected.
 
Photon candidates are reconstructed from clusters of energy deposits in the EM calorimeter, calibrated at the EM scale, and tracking information from the ID, which is used to classify candidates as either converted or unconverted photons. Candidate photons are required to have a transverse momentum $p_{\mathrm{T}}>20$~GeV and a pseudorapidity of $|\eta|<2.37$, excluding the transition region between the electromagnetic barrel and endcap regions of the calorimeter, $1.37<|\eta|<1.52$. Photons must also satisfy the cut-based \textit{Tight} identification requirement defined using EM shower shape variables~\cite{PERF-2017-02}. To reject non-prompt photons originating from jets, photons must satisfy an isolation requirement based on topological clusters~\cite{PERF-2014-07} of energy deposits in the EM calorimeter. The isolation energy of a photon, $\etisoy$, is determined by first calculating the scalar sum of the transverse energy of topological clusters within $\Delta R = 0.4$ of a photon ($\etcone$), corrected for the energy of the photon itself, and then subtracting off a value that depends on the transverse photon energy ($\ET^{\gamma}$), such that $\etisoy = \etcone - 0.022*\ET^{\gamma}$. Photons are required to pass the \textit{Calorimeter-Only \Tight} isolation working point~\cite{PERF-2017-02}, which requires $\etisoy < 2.45$~GeV. In addition, the two photons must be  separated from each other by requiring $\Delta R >0.4$.
 
Electron candidates are reconstructed from energy deposits in the EM calorimeter that can be matched to ID tracks. These tracks must be consistent with originating from the primary vertex by requiring that
$|d_0 / \sigma_{d_0}|<5$ and $|z_0\cdot\mathrm{sin}(\theta)|<0.5$~mm, where $d_0$ is the transverse impact parameter relative to the beam line and $\sigma_{d_0}$ is its uncertainty, $z_0$ is the longitudinal impact parameter, and $\theta$ is the polar angle of the track with respect to the beamline. Electron candidates are required to have $p_{\mathrm{T}}>25$~GeV and $|\eta|<2.47$, excluding the region $1.37<|\eta|<1.52$. Additionally, they must satisfy the likelihood-based \Medium identification requirement defined using inputs from the calorimetry and tracking systems~\cite{PERF-2017-01}.
 
Muon candidates are reconstructed by matching tracks in the ID to tracks in the MS. These tracks must be consistent with originating from primary vertices by requiring $|d_0 / \sigma_{d_0}|<3$ and $|z_0\cdot\mathrm{sin}(\theta)|<0.5$~mm. Muon candidates are further required to have $p_{\mathrm{T}}>25$~GeV and $|\eta|<2.4$, and must satisfy the \Medium identification requirement~\cite{PERF-2015-10} based on the quality and compatibility of their tracks in the ID and MS.
 
To further distinguish signal leptons from background, isolation variables for calorimeter energy deposits ($\etiso$) and tracks ($\ptiso$) are constructed. The calorimeter isolation $\etiso$ is defined as the scalar sum of the transverse energy of topological clusters within $\Delta R = 0.2$ of the lepton, which is corrected for both the energy of the lepton itself and the average pileup energy density measured in this region of the detector. For electrons (muons), the track-based isolation $\ptisovare$ ($\ptisovarmu$) is defined as the scalar sum of tracks with $\pT > 1$~GeV within a $\pT$-dependent cone up to $\Delta R = 0.2$ ($\Delta R = 0.3$) of the lepton, with the lepton candidate removed. Electrons must satisfy $\etiso/\pT < 0.06$ and $\ptisovare/\pT < 0.06$~\cite{PERF-2017-01}, and muons must satisfy $\etiso/\pT < 0.15$ and $\ptisovarmu/\pT < 0.04$~\cite{MUON-2018-03}.
 
Hadronic jets are used in the SR definition to veto events with jets containing $b$-hadrons. Jets are reconstructed using the \antikt algorithm~\cite{antikt1} with a distance parameter $\Delta R = 0.4$.  The inputs to the jet algorithm are particle-flow objects~\cite{PERF-2015-09}, which make use of both
the calorimeter and the ID information to precisely determine the momenta of the input
particles.
Reconstructed jets are required to have $p_{\mathrm{T}}>20$~GeV and $|\eta|<4.5$. Jets satisfying $20<p_{\mathrm{T}}<60$~GeV and $|\eta|<2.4$ must pass the \Tight requirement on the \textit{jet vertex tagger} variable~\cite{PERF-2014-03} in order to suppress jets not originating from the primary vertex.  Kinematic properties of $b$-flavored hadrons are used as input to a multivariate jet classification algorithm~\cite{FTAG-2018-01, PERF-2012-04}. This multivariate classification has a 77\% efficiency and
is used to identify jets with $|\eta|<2.5$ containing $b$-flavored hadrons. Events with jets containing $b$-flavored hadrons are rejected.
 
It is possible for tracks and energy deposits to be associated with more than one type of reconstructed object.
To remove the overlap between different reconstructed objects in an event, the following selection criteria are applied in the order in which they are described:  electrons are removed if they share an ID track with a muon; photons are removed if they are within $\Delta R = 0.4$ of an electron or a muon; jets are removed if they are within $\Delta R = 0.2$ of an electron; electrons are removed if they are within $\Delta R = 0.4$ of a jet; jets are removed if they are within $\Delta R = 0.2$ of a muon; and finally photons and muons are removed if they are within $\Delta R = 0.4$ of the remaining jets in the event.
 
The magnitude and direction of the missing transverse momentum are reconstructed using calibrated photons, electrons, muons, jets, and tracks from charged particles not associated to any object found in the event~\cite{ATLAS-CONF-2018-023}. An ambiguity resolution procedure is performed as part of the calculation to ensure that energy deposits reconstructed as different objects are not double-counted. Events in the SR are required to satisfy $\MET>25$~GeV. Additionally, the transverse mass of the $W$ boson $m^{W}_{\mathrm{T}} = \sqrt{2 p^{\ell}_{\mbox{\footnotesize{T}}} \MET (1-\cos{\Delta \phi})}$ is required to be greater than $40$~GeV, where $\Delta \phi$ is defined as the difference in azimuthal angles between the lepton momentum and missing transverse momentum.
 
A set of $Z\gamma$ veto requirements are implemented to greatly reduce the number of $Z\gamma$ events passing the signal selection, which can occur when an electron is misidentified as a photon. All SR events must have $p_{\mathrm{T},\ell\gamma\gamma}>30$~GeV and $m_{\ell\gamma\gamma}$, $m_{\ell\gamma_1}$, and $m_{\ell\gamma_2} \notin [82, 100]$~GeV, where $\gamma_1$ and $\gamma_2$ are the leading and sub-leading photons ordered by $p_{\mathrm{T}}$. The $Z\gamma$ veto is applied to both the electron and muon channels to ensure a consistent event selection.
 
To reduce contributions from background events with a second lepton originating from processes with two $W$ bosons or a $Z$ boson, two additional selection criteria are applied.
Events are rejected if they contain a second lepton, selected without the $|d_0/\sigma_{d_0}|$ or isolation requirements, of a different lepton flavor to the primary lepton that passes all SR selection criteria.
A similar veto is enforced for events containing same-flavor leptons with the secondary lepton only required to satisfy $p_{\mathrm{T}}>6$~GeV and pass \Loose(\Medium) identification for electrons~\cite{PERF-2017-01} (muons~\cite{PERF-2015-10}), where the lepton identification requirement is loosened to remove a significant fraction of the prompt background from the $Z\gamma\gamma$ process.
 
Differences in the reconstruction, trigger, and selection efficiencies for leptons and photons between data and simulation are corrected for with scale factors~\cite{EGAM-2018-01,TRIG-2018-05,,MUON-2018-03,TRIG-2018-01}. In addition to the SR defined in this section, other data samples are used to estimate backgrounds coming from misidentified objects using data-driven techniques, as described in Section~\ref{sec:background}.
 
\section{Background Estimation}
\label{sec:background}
 
The largest background in the $W\gamma\gamma$ SR consists of events in which one or both signal photons originate from a misidentified jet or neutral hadron decay.
This hadronic fake photon background, denoted as $j \rightarrow \gamma$, is estimated using a data-driven method by performing a two-dimensional template fit to the isolation distributions of the leading and subleading photons in a procedure similar to those discussed in Refs.~\cite{STDM-2013-05}~and~\cite{STDM-2011-05}.
The three isolation distribution templates for the cases in which either the leading, subleading, or both photons are $j \rightarrow \gamma$ fakes are obtained from data in regions formed by loosening and inverting some of the leading, subleading, or both photon isolation requirements, respectively, in order to enhance the contributions from misidentified jets.
This is done by selecting events in which at least one photon candidate passes the \Loose photon identification requirement but fails one or more of the four EM shower-shape requirements used in the \Tight (T) photon identification~\cite{PERF-2017-02}; these are denoted as L$'$ photons~\cite{EGAM-2018-01}.
For the estimation of this source of background, events are still required to satisfy all other SR criteria except the photon isolation requirement.
The electron and muon channels are combined to ensure a sufficient number of events pass selection requirements for the data samples.
These events are categorized into four non-overlapping data samples, TT, TL$'$, L$'$T, and L$'$L$'$, depending whether the T or L$'$ photon identification criteria is satisfied by the leading and subleading photons, respectively.
Templates for non-prompt leading and subleading photons are built using one-dimensional Bukin functions~\cite{Bukin:2007zha}, and their shape parameters are determined from fits to photon isolation energy distributions of data events in the L$'$T and TL$'$ regions, respectively.
The templates for leading (subleading) prompt photons are formed from double-sided crystal ball functions, whose shape parameters are fit to simulated tight leading (subleading) prompt photons in simulated $W\gamma\gamma$ events; these correspond to leading photons from events in the TT and TL$'$ regions and subleading photons from events in the TT and L$'$T regions.
Two-dimensional templates for prompt $\gamma\gamma$, $\gamma(j\rightarrow\gamma)$, and $(j\rightarrow\gamma)\gamma$ events are formed by taking the product of the two functions used to individually describe the isolation energy of the leading and subleading photons.
Due to non-negligible correlations between the two photon candidates in the L$'$L$'$ data sample, the two-dimensional template for $(j\rightarrow\gamma)(j\rightarrow\gamma)$ events is instead formed by fitting a superposition of Gaussian kernels~\cite{Cranmer:2000du}.
Finally, coefficients corresponding to numbers of events for each of the four, two-dimensional templates are fit using an extended maximum likelihood fit to data in the TT region that has simulated events from all other background processes with two prompt photons and one prompt lepton subtracted.
The coefficients for the TL$'$, L$'$T, and L$'$L$'$ regions are further corrected for signal leakage using MC simulation.
In order to account for the photon isolation energy requirements that are part of the SR definition, the contribution from $j\rightarrow\gamma$ fake events in the SR is obtained by integrating the 2D photon isolation energy distributions fitted to data in the TT region up to the cut value that defines the SR, $\etisoy= 2.45$~GeV.
The total expected number of $j\rightarrow\gamma$ background events is determined by computing the sum of the integrated coefficients of the $\gamma(j\rightarrow\gamma)$, $(j\rightarrow\gamma)\gamma$, and $(j\rightarrow\gamma)(j\rightarrow\gamma)$ templates.
A systematic uncertainty due to the choice of photon $L'$ identification is estimated by forming $j\rightarrow\gamma$ templates with alternative identification working points and parameterizing the shape differences as uncertainties on the nominal Bukin template parameters. Statistical and systematic uncertainties on the templates are propagated through to the background estimates using a multivariate Gaussian constraint on the two-dimensional fit, resulting in an overall 11\% systematic uncertainty on this background.
 
Events in which one or both photons are misidentified electrons constitute the $e\rightarrow\gamma$ fake background.
These misidentifications are caused mainly by tracking inefficiencies and the mismatching of tracks in the ID to energy clusters in the EM calorimeter. The background is estimated using a data-driven ``fake-factor'' method similar to the one described in Ref.~\cite{PERF-2017-02}. The $e\rightarrow\gamma$ fake rate is calculated with a tag-and-probe approach using both $Z\rightarrow ee$ and $Z\rightarrow e\gamma_{e}$ events, where $\gamma_{e}$ symbolizes a misreconstructed electron identified as a photon. Probe electrons are selected with $\pT > 20$~GeV, the likelihood-based \Tight identification~\cite{PERF-2017-01}, and the same isolation requirement as SR electrons, such that their kinematics selection is close to the one for the photons used in the SR. The data sample used to calculate the $e\rightarrow\gamma$ fake rate consists of events selected with single electron triggers that have a reconstructed $ee/e\gamma_{e}$ invariant mass within 20~GeV of the $Z$ boson mass, 91.2~GeV.
Non-resonant backgrounds in this data sample are modeled using an exponential function, and the $Z$ boson resonance is modeled by a Gaussian with double-sided exponential tails~\cite{GausDoubleExp}. The number
of $ee$ ($N_{ee}$) and $e\gamma_{e}$ ($N_{e\gamma_{e}}$) events are extracted using a combined signal and background fit to the invariant mass distributions in bins of \pT and $\eta$, and the fake factor is computed as $f_{e\rightarrow\gamma_{e}} = N_{e\gamma_{e}}/ N_{ee}$.  To estimate the $e\rightarrow\gamma$ background in the SR, this fake factor is applied to a sample of $W(\ell\nu)e\gamma$ events obtained by selecting data events with di-lepton triggers
and substituting one SR photon requirement for that of a probe electron.
Systematic uncertainties relating to the fitting and integration ranges around the $Z$-boson mass, the photon energy calibration, and the exponential background are propagated to the $e\rightarrow\gamma$ background estimate. Statistical and systematic uncertainties are 2\% and 7\%, respectively, on the final SR $e\rightarrow\gamma$ background estimate. A validation region dominated by events with $e\rightarrow\gamma$ fakes is obtained by inverting the $Z\gamma$-veto in the SR and in the  $W(\ell\nu)e\gamma$ region.  The background estimation method is shown to reproduce data in both of these validation regions.
 
The hadronic fake lepton background, $j\rightarrow\ell$, is comprised of events in which the signal lepton is either a misidentified jet or from the decay of a heavy-flavored hadron (non-prompt). This background is also estimated using a data-driven fake-factor~\cite{EGAM-2019-01} using an event sample that is enriched in non-prompt leptons.  This data sample is obtained  by selecting $Z(\ell\ell)+\ell_{j}$ events with single lepton triggers
where the third lepton, $\ell_j$, is a misreconstructed jet. Two leptons must have an invariant mass within 10\% of the $Z$-boson mass and be of the same flavor but opposite charge, while the third (probe) lepton must be of a different lepton flavor in order to avoid ambiguity. Additional requirements of $\MET < 40$~GeV and $m^{W}_{\mathrm{T}} < 40$~GeV are imposed to reduce prompt leptons from $WZ$ events, and remaining $WZ$ events are subtracted from the data, relying on simulated predictions.
The fake factor is defined as the ratio of the number of probe leptons satisfying the SR lepton criteria ($N_{\mbox{\footnotesize{SR}}}$), to the number of probe leptons satisfying a \Loose set of criteria ($N_{l}$). This \Loose criteria selects leptons more likely to be non-prompt by inverting the lepton $|d_0/\sigma_{d_0}|$ and isolation requirements. The fake factor is estimated in bins of probe lepton \pT and $|\eta|$ for electrons, and only in bins of \pT for muons due to statistical limitations. The $j\rightarrow\ell$ fake background in the SR is estimated by applying the fake factor to a region kinematically adjacent to the SR. This region is defined with the same selection requirements as for the SR with the exception of the lepton selection, which uses the \Loose selection criteria. Statistical and systematic uncertainties account for a 26\% (27\%) and 18\% (50\%) uncertainty in the electron (muon) channel, respectively. Systematic uncertainties relating to a bias in the control region due to the \MET selection are computed by varying the requirement by $\pm10$~GeV~\cite{EGAM-2019-01}, and theoretical uncertainties on the subtracted $WZ$ events are propagated through to the fake factors. The method is validated in a region enriched in fake leptons obtained by inverting the  \MET  and $m^{W}_{\mathrm{T}}$ requirements used in the SR and comparing the estimate to data.
 
The pileup background consists of events in which one or both photons do not originate from the primary vertex, mainly due to a limited photon pointing resolution.
The fraction of photons originating from a pileup vertex is calculated in a subset of SR data where at least one photon is converted. Since the fraction of photons that convert is independent of their production vertex, the relative fractions of signal and pileup photons in the converted sample is representative of the fractional number of signal and pileup photons in the full SR.
Converted photons that are required to have at least one ID track with silicon hits~\cite{EGAM-2018-01} and a conversion radius, defined as the radial distance of the conversion vertex, of less than 400~mm are used for this estimate because the presence of an ID track allows for the calculation of a longitudinal impact parameter.
The difference between the longitudinal impact parameters of the converted photon and the primary vertex, $\Delta z$, is Gaussian-distributed and expected to be close to zero for photons from the hard scatter, while pileup photons are expected to have a much broader distribution~\cite{STDM-2018-04}.
The $\left|\Delta z\right| > 55$~mm tails of the distribution are used to estimate the fraction of pileup photons in the SR. The statistical uncertainty on the pileup background is $56\%$, due to the limited number of events in the estimation region.
 
Event yields in the SR from irreducible sources of background such as $WH(\gamma\gamma)$, $WW\gamma$, and $Z\gamma\gamma$ as well as $tt\gamma$, $tW\gamma$, and $tq\gamma$ are estimated using MC simulated samples. To further reduce uncertainties from the estimated $tt\gamma$ event yield in the SR, a control region enhanced in $tt\gamma$ events (TopCR) is defined by inverting the $b$-jet veto in the SR selection requirements in order to constrain a $tt\gamma$ normalization factor that is left floating in the likelihood fit described in Section~\ref{sec:results}. The fitted $tt\gamma$ normalization factor is cross-checked in a validation region (TopVR) formed by inverting the $b$-jet veto, the \MET, and $m^{W}_{\mathrm{T}}$ in the SR selection requirements in order to select events with at least one $b$-jet, $\MET < 25$~GeV, and $m^{W}_{\mathrm{T}} < 40$~GeV.
The $j\rightarrow\gamma$ and $e\rightarrow\gamma$ data-driven backgrounds are also computed in the TopCR and TopVR following the same methods outlined for the SR. Due to the reduced number of events in the L$'$ regions, the photon identification systematic uncertainty is estimated using a dedicated procedure in the SR, and is $+18\%/-13\%$. The $j\rightarrow\ell$ and pileup backgrounds are negligible in both of these TopCR and TopVR regions.
 
\section{Uncertainties}
 
The background uncertainties described in Section~\ref{sec:background} are the dominant uncertainties of this measurement described in Section~\ref{sec:results}. In addition, several other important sources of uncertainty are assessed. These include instrumental uncertainties such as the energy scale and resolution of electrons and photons~\cite{EGAM-2018-01}; photon and lepton trigger, reconstruction, identification, and isolation efficiencies~\cite{TRIG-2018-01, TRIG-2018-05, EGAM-2018-01, PERF-2015-10}; jet energy scale and resolution~\cite{JETM-2018-05}; jet vertex tagging~\cite{ATL-PHYS-PUB-2014-001, PERF-2016-06}; $b$-jet identification~\cite{FTAG-2018-01}; missing transverse energy reconstruction~\cite{PERF-2016-07}; and the luminosity of the dataset~\cite{DAPR-2021-01}. These are evaluated for both backgrounds and signal processes.
 
Additionally, theoretical uncertainties associated with the simulation of the signal and background processes are evaluated and propagated through to the measured fiducial cross section. Theoretical uncertainties on the background processes, but not signal processes, are propagated through to the measured signal strength. These include parton distribution function uncertainties~\cite{Butterworth:2015oua}; the uncertainty on the strong coupling constant, $\alpha_{\mathrm{s}}$~\cite{Martin:2009bu}; and missing higher-order terms in the cross section calculations~\cite{ATL-PHYS-PUB-2016-002}. The last is evaluated by varying the renormalization and factorization scales independently by factors of 0.5 and 2, avoiding variations where the two scales differ by more than a factor of two.
 
Statistical uncertainties on the data, and signal and background MC samples are also taken into account. All of the previously described uncertainties are accounted for in the detector-to-fiducial region correction factor used for the unfolding procedure detailed in Section~\ref{sec:results}.
 
\section{Results}
\label{sec:results}
 
The $pp \rightarrow W\gamma\gamma$ signal strength $\mu$ is extracted from the data using a binned maximum likelihood fit~\cite{Cranmer:2012sba, James:1975dr} including the TopCR and the signal region. All uncertainties considered in the analysis are treated as nuisance parameters in the fit. Systematic uncertainties are constrained by Gaussian functions, and correlations between sources of systematic uncertainties are taken into account. Statistical uncertainties are also treated as nuisance parameters but are constrained by assigning a Poisson function to each analysis bin. These constraints penalize the likelihood fit if the estimated nuisance parameters pull from their measured values.
 
\begin{figure}[ht!]
\centering
\includegraphics[width=\textwidth]{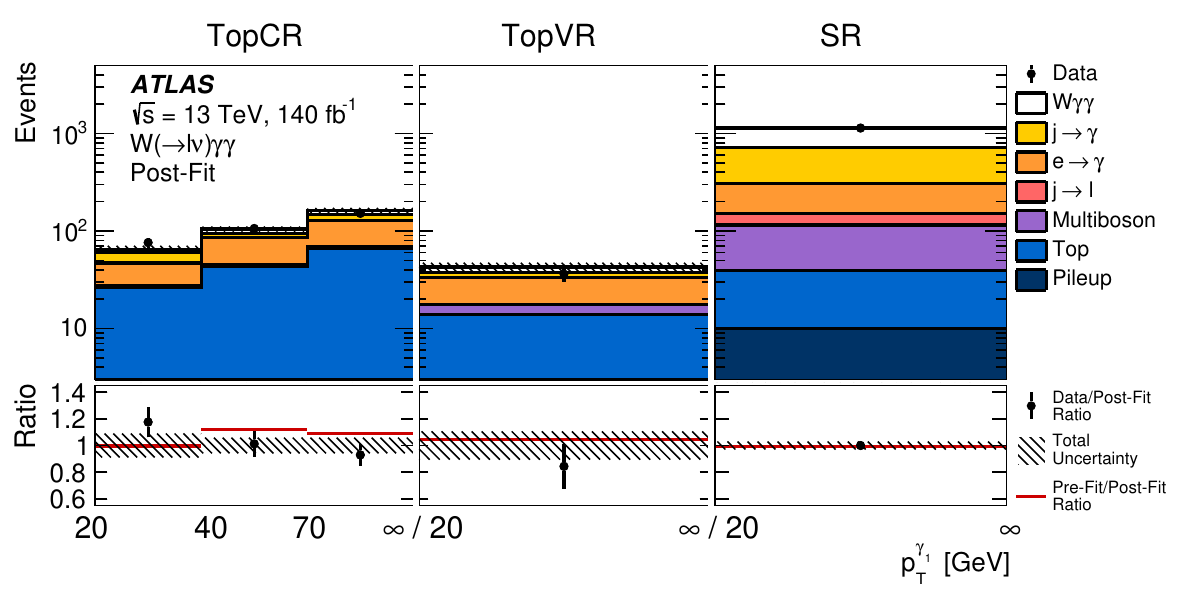}
\caption{Data, and pre- and post-fit yields for TopCR as a function of leading photon \pT, and for TopVR and SR each as a single bin. The error bars on data indicate its statistical uncertainty. The bottom panel shows the ratio of the data to the post-fit yield (black points) and the ratio of the pre-fit yield to the post-fit yield (solid line) for each of the regions. The uncertainty band includes both the statistical and systematic uncertainties obtained from the fit. The $tt\gamma$ background is scaled by the normalization factor $\alpha_{tt\gamma}$, and the $W(\ell\nu)\gamma\gamma$ prediction, by the signal strength $\mu$. Background contributions from pileup in TopCR and TopVR are neglected.}
\label{fig:result}
\end{figure}

The TopCR is used to determine a $tt\gamma$ background\footnote{Background from top processes $tW\gamma$ and $tq\gamma$ are small and are thus not included in the determination and application of $\alpha_{tt\gamma}$.} normalization factor $\alpha_{tt\gamma}$. The normalization factor is allowed to float via a likelihood scan done simultaneously with the signal-strength extraction in the SR. The fit value of $\alpha_{tt\gamma}$ is then applied in the TopVR and the resulting total estimated yield is compared to data.
Figure \ref{fig:result} illustrates the yields for the
three regions TopCR, TopVR, and SR.  The estimated yield in the TopVR region shows agreement with data.
 
\begin{table}[ht!]
\caption{Estimated signal and background yields  in the SR and TopCR, as well as their sums, are shown post-fit together with the observed number of events in data. The uncertainties quoted in the table correspond to total uncertainties. Events from the ``Multiboson'' and ``Top'' backgrounds are estimated from MC simulation and contain only prompt leptons and photons. Yields denoted with ``--'' correspond to backgrounds that are negligible.}
\centering
\sisetup{group-minimum-digits=4}
 
\begin{tabular}{lr@{$\,\pm\,$}lr@{$\,\pm\,$}l}
\toprule
Source                                        & \multicolumn{2}{c}{SR} & \multicolumn{2}{c}{TopCR} \\
\midrule\midrule
$W\gamma\gamma$                          & 410& 60 & 28 & 5\\
Non-prompt $j \rightarrow \gamma$     & 420 & 50 & 42& 20 \\
Misidentified $e \rightarrow \gamma$              & 155 & 11 & 120 & 9\\
Multiboson ($WH(\gamma\gamma)$, $WW\gamma$, $Z\gamma\gamma$)  & 76 & 13 & 5.2 & 1.7\\
Non-prompt $j \rightarrow \ell$                & 35 & 10 &  \multicolumn{2}{c}{--} \\
Top ($tt\gamma$, $tW\gamma$, $tq\gamma$)  & 30 & 7 & 136 &32\\
Pileup                                   & 10 &5 &  \multicolumn{2}{c}{--} \\
\midrule
Total                                    & \num{1136} &34 & 332 & 18\\
Data                                    & \multicolumn{2}{c}{\num{1136}} & \multicolumn{2}{c}{333} \\
\bottomrule
\end{tabular}
\label{tab:yields}
\end{table}
 
Table \ref{tab:yields} shows the post-fit yields of the signal and estimated backgrounds in the SR and TopCR, along with their sum and the number of selected data events.
The signal strength and $tt\gamma$ normalization are determined to be $\mu = 1.01\substack{+0.17 \\ -0.16}$ and $\alpha_{tt\gamma} = 0.83\substack{+0.21 \\ -0.25}$ and have been applied to the post-fit results. Assuming lepton universality and correct modeling of $\tau$-to-$e$ and $\tau$-to-$\mu$ decays, events from $W(\tau\nu)\gamma\gamma$ with leptonic $\tau$ decays that fall into the fit regions are included as a part of the signal in the fitting procedure and are normalized together with $W(e\nu)\gamma\gamma$ and $W(\mu\nu)\gamma\gamma$. The fit results yield an expected and observed significance of 5.6 standard deviations, corresponding to the observation of the $W\gamma\gamma$ process. No nuisance parameters are significantly pulled or constrained in the fit.

\begin{figure}[!ht]
\centering
\includegraphics[width=0.6\textwidth]{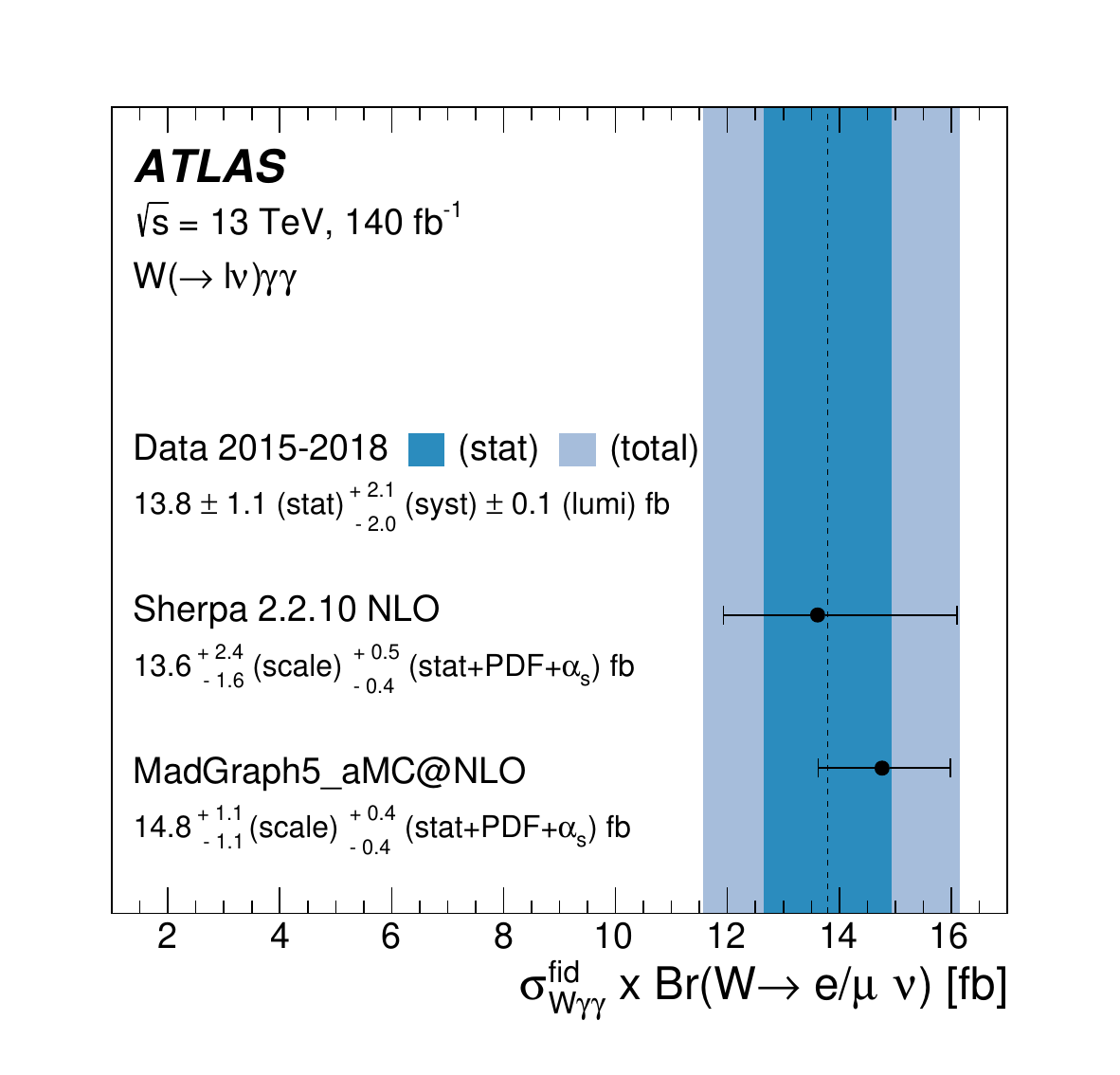}
\caption{The measured fiducial $W(\rightarrow e\nu / \mu\nu)\gamma\gamma$ integrated cross section compared with both the signal event generator predictions.}
\label{fig:xsec}
\end{figure}
 
\begin{table}[ht!]
\caption{Major sources of uncertainty and their impacts on the measured fiducial cross section, as calculated from the correlation matrix of the fiducial cross section fit.
Squared values of impacts are determined by setting all nuisance parameters for a given uncertainty source to their best-fit value and subtracting the resulting squared value of the total uncertainty from the squared value of the total uncertainty in the nominal fit.
Systematic uncertainty sources that contribute less than 0.1\% are not shown. Efficiency uncertainties include, where applicable, uncertainties on data-MC agreement due to reconstruction, trigger selection, identification, isolation, and vertex-matching.}
\centering
\begin{tabular}{lc}
\toprule
Source of uncertainty           & Impact [\%] \\
\midrule
\midrule
$j\rightarrow\gamma$ data-driven background estimate     	   				& 12\\
Photon efficiency 					& 4.5\\
Other data-driven background estimates     	   				& 3.5\\
Background MC theoretical modeling   							& 3.0\\
Monte Carlo statistics						& 2.7\\
Signal MC theoretical modeling                    & 2.6\\
Jet efficiency and calibration			   					& 2.4\\
Top normalization  				& 2.3\\
Pileup reweighting					& 1.6\\
Muon efficiency	and calibration		  					& 1.4\\
\MET calibration  					& 1.3\\
Luminosity  						& 1.0\\
Electron and photon calibration	   				& 0.7\\
Flavor tagging efficiency 			& 0.6\\
\midrule
Systematic  						& 15\\
Statistical							& 8.3\\
\midrule
Total	    						& 17\\
\bottomrule
\end{tabular}
\label{tab:systematics}
\end{table}
 
In order to obtain an unfolded production cross section measurement, a fiducial phase space is defined to be as close as possible to the SR event sample selected at detector-level. Fiducial requirements are applied to dressed leptons, which are particle-level electrons and muons recombined with radiated photons within a cone of $\Delta R = 0.1$. Events are required to have a dressed electron or muon with $p_{\mathrm{T}} > 25$~GeV and $|\eta| < 2.47$ while the two particle-level photons must satisfy $p_{\mathrm{T}} > 20$~GeV and $|\eta| < 2.37$. Additionally, photons must satisfy the isolation requirement ($\etconegen - 0.032 \times \ET) < 6.53$~GeV, where $\etconegen$ is computed from the vector momentum sum of all stable, generator-level particles within $\Delta R = 0.4$ of the photon. This isolation requirement is derived to vary with photon $\ET$ to mimic the detector-level isolation requirement. Additionally, two separation requirements are applied to the two photons and between the lepton and each photon: $\Delta R_{\gamma\gamma} > 0.4$ and $\Delta R_{\ell\gamma} > 0.4$. Finally, fiducial events must satisfy $E^{\mathrm{miss}}_{\mathrm{T}} > 25$~GeV, $m^{W}_{\mathrm{T}} > 40$~GeV, and a veto on $b$-jets with $\pT > 20$~GeV and $|\eta| < 2.5$. The unfolding is performed into a fiducial phase space with $W\rightarrow\mu\nu$ and $W\rightarrow e\nu$ decays;  $W\rightarrow\tau\nu$ decays that pass these requirements, including events in which the tau decays leptonically, are not considered as part of the fiducial phase space.
 
Unfolding is performed on the measurement using a similar maximum likelihood method to the one used to perform the signal-strength extraction, where the effects of statistical, experimental, and theoretical uncertainties on the modeling of the correction from detector-level signal events to the fiducial phase space are taken into account.
A correction factor $C$ is calculated as the ratio of the number of signal MC events reconstructed in the signal region to the predicted in the fiducial phase-space.
The number of detector-level events is defined as the sum of simulated MC events with two photons and $W$-boson decaying into an electron, muon, or leptonically decaying tau that pass all signal region requirements.
The number of fiducial events is calculated using only simulated $W(e\nu)\gamma\gamma$ and $W(\mu\nu)\gamma\gamma$ signal MC events, where the electron or muon is prompt.
The correction factor is computed to be $C = 0.210 \pm 0.004 (\mathrm{stat.})$ using the \textsc{Sherpa} NLO signal MC samples; a 2.9\% relative difference is found when calculating $C$ with \textsc{MadGraph} signal MC samples, which is in statistical agreement and thus no generator choice uncertainty is added. In the likelihood fit, the total number of expected signal events is defined as
$N_{\mathrm{sig}} = C\,\sigma_{\mathrm{pred}}^{e/\mu}\mathcal{L}$.
The signal production cross section is measured in the fiducial phase space from the number of signal events observed in data, the integrated luminosity, and the correction factor.
The measured fiducial cross section for $W(e/\mu\nu)\gamma\gamma$ events is determined to be $\sigma_{\mathrm{fid}} = 13.8 \pm 1.1 (\mathrm{stat}) \substack{+2.1 \\ -2.0} (\mathrm{syst}) \pm 0.1 (\mathrm{lumi})$~fb and it is in close agreement with SM predictions as shown in Figure~\ref{fig:xsec}.

In Table \ref{tab:systematics}, the dominant sources of uncertainty and their impact on the fiducial cross section are listed. For the purposes of this table, the uncertainties are grouped into common categories given their source.
The impact of each group of systematic uncertainties is calculated by performing the likelihood fit where the individual parameters of the grouped systematics are set to their best fit values from the nominal fit and not allowed to float.
For each grouping, the square value of the new overall fit uncertainty is subtracted from the squared value of the nominal fit uncertainty to obtain the squared value of the impact of the grouped uncertainties.
The fit is performed under the assumption that the nuisance parameters for the grouped systematics that are held fixed are uncorrelated to all others that are allowed to float.
This procedure is used only to estimate the impact of the individual groups of systematics, as it avoids the possibility of abnormal pulls that could occur if the fit were performed with only one group of nuisance parameters left floating at a time.
The largest source of systematic uncertainty is due to the $j\rightarrow\gamma$ data-driven background estimate, followed by the statistical uncertainty on data. The modeling of  the identification, isolation, and trigger efficiencies to select photons in simulated $W\gamma\gamma$ events also represents a substantial source of uncertainty, and together these comprise the ``Photon efficiency'' uncertainty source in Table~\ref{tab:systematics}.
 
\FloatBarrier
 
\section{Conclusion}
 
This letter reports the observation and measurement of the process $pp \rightarrow W(\ell\nu)\gamma\gamma$ by the ATLAS experiment at the LHC. Leptonic decays of the $W$ boson to an electron or a muon accompanied by two photons are selected from the 140~fb$^{-1}$ \RunTwo dataset of proton-proton collisions at $\sqrt{s} = 13$~TeV produced by the LHC. A maximum likelihood fit of the signal and background yields leads to a rejection of the background-only hypothesis with an observed and expected significance of $5.6$ standard deviations. The measured fiducial cross section for $W(e\nu)\gamma\gamma$ and $W(\mu\nu)\gamma\gamma$ events  is $\sigma_{\mathrm{fid}} = 13.8 \pm 1.1 (\mathrm{stat}) \substack{+2.1 \\ -2.0} (\mathrm{syst}) \pm 0.1 (\mathrm{lumi})$~fb, in agreement with the SM predictions for this process. The dominant sources of uncertainty come from the data-driven background estimates and the statistical uncertainty on data.
 
\section*{Acknowledgements}


We thank CERN for the very successful operation of the LHC, as well as the
support staff from our institutions without whom ATLAS could not be
operated efficiently.
 
We acknowledge the support of
ANPCyT, Argentina;
YerPhI, Armenia;
ARC, Australia;
BMWFW and FWF, Austria;
ANAS, Azerbaijan;
CNPq and FAPESP, Brazil;
NSERC, NRC and CFI, Canada;
CERN;
ANID, Chile;
CAS, MOST and NSFC, China;
Minciencias, Colombia;
MEYS CR, Czech Republic;
DNRF and DNSRC, Denmark;
IN2P3-CNRS and CEA-DRF/IRFU, France;
SRNSFG, Georgia;
BMBF, HGF and MPG, Germany;
GSRI, Greece;
RGC and Hong Kong SAR, China;
ISF and Benoziyo Center, Israel;
INFN, Italy;
MEXT and JSPS, Japan;
CNRST, Morocco;
NWO, Netherlands;
RCN, Norway;
MEiN, Poland;
FCT, Portugal;
MNE/IFA, Romania;
MESTD, Serbia;
MSSR, Slovakia;
ARRS and MIZ\v{S}, Slovenia;
DSI/NRF, South Africa;
MICINN, Spain;
SRC and Wallenberg Foundation, Sweden;
SERI, SNSF and Cantons of Bern and Geneva, Switzerland;
MOST, Taiwan;
TENMAK, T\"urkiye;
STFC, United Kingdom;
DOE and NSF, United States of America.
In addition, individual groups and members have received support from
BCKDF, CANARIE, Compute Canada and CRC, Canada;
PRIMUS 21/SCI/017 and UNCE SCI/013, Czech Republic;
COST, ERC, ERDF, Horizon 2020 and Marie Sk{\l}odowska-Curie Actions, European Union;
Investissements d'Avenir Labex, Investissements d'Avenir Idex and ANR, France;
DFG and AvH Foundation, Germany;
Herakleitos, Thales and Aristeia programmes co-financed by EU-ESF and the Greek NSRF, Greece;
BSF-NSF and MINERVA, Israel;
Norwegian Financial Mechanism 2014-2021, Norway;
NCN and NAWA, Poland;
La Caixa Banking Foundation, CERCA Programme Generalitat de Catalunya and PROMETEO and GenT Programmes Generalitat Valenciana, Spain;
G\"{o}ran Gustafssons Stiftelse, Sweden;
The Royal Society and Leverhulme Trust, United Kingdom.
 
The crucial computing support from all WLCG partners is acknowledged gratefully, in particular from CERN, the ATLAS Tier-1 facilities at TRIUMF (Canada), NDGF (Denmark, Norway, Sweden), CC-IN2P3 (France), KIT/GridKA (Germany), INFN-CNAF (Italy), NL-T1 (Netherlands), PIC (Spain), ASGC (Taiwan), RAL (UK) and BNL (USA), the Tier-2 facilities worldwide and large non-WLCG resource providers. Major contributors of computing resources are listed in Ref.~\cite{ATL-SOFT-PUB-2021-003}.


\printbibliography
 
\clearpage
 
\begin{flushleft}
\hypersetup{urlcolor=black}
{\Large The ATLAS Collaboration}

\bigskip

\AtlasOrcid[0000-0002-6665-4934]{G.~Aad}$^\textrm{\scriptsize 103}$,
\AtlasOrcid[0000-0002-5888-2734]{B.~Abbott}$^\textrm{\scriptsize 121}$,
\AtlasOrcid[0000-0002-1002-1652]{K.~Abeling}$^\textrm{\scriptsize 55}$,
\AtlasOrcid[0000-0001-5763-2760]{N.J.~Abicht}$^\textrm{\scriptsize 49}$,
\AtlasOrcid[0000-0002-8496-9294]{S.H.~Abidi}$^\textrm{\scriptsize 29}$,
\AtlasOrcid[0000-0002-9987-2292]{A.~Aboulhorma}$^\textrm{\scriptsize 35e}$,
\AtlasOrcid[0000-0001-5329-6640]{H.~Abramowicz}$^\textrm{\scriptsize 152}$,
\AtlasOrcid[0000-0002-1599-2896]{H.~Abreu}$^\textrm{\scriptsize 151}$,
\AtlasOrcid[0000-0003-0403-3697]{Y.~Abulaiti}$^\textrm{\scriptsize 118}$,
\AtlasOrcid[0000-0003-0762-7204]{A.C.~Abusleme~Hoffman}$^\textrm{\scriptsize 138a}$
% ...
\bigskip
\\

% ...
$^{29}$Physics Department, Brookhaven National Laboratory, Upton NY; United States of America.\\
% ...
$^{35}$$^{(a)}$Facult\'e des Sciences Ain Chock, R\'eseau Universitaire de Physique des Hautes Energies - Universit\'e Hassan II, Casablanca;$^{(b)}$Facult\'{e} des Sciences, Universit\'{e} Ibn-Tofail, K\'{e}nitra;$^{(c)}$Facult\'e des Sciences Semlalia, Universit\'e Cadi Ayyad, LPHEA-Marrakech;$^{(d)}$LPMR, Facult\'e des Sciences, Universit\'e Mohamed Premier, Oujda;$^{(e)}$Facult\'e des sciences, Universit\'e Mohammed V, Rabat;$^{(f)}$Institute of Applied Physics, Mohammed VI Polytechnic University, Ben Guerir; Morocco.\\
% ...
$^{49}$Fakult\"{a}t Physik , Technische Universit{\"a}t Dortmund, Dortmund; Germany.\\
% ...
$^{55}$II. Physikalisches Institut, Georg-August-Universit\"{a}t G\"ottingen, G\"ottingen; Germany.\\
% ...
$^{103}$CPPM, Aix-Marseille Universit\'e, CNRS/IN2P3, Marseille; France.\\
% ...
$^{118}$Department of Physics, New York University, New York NY; United States of America.\\
% ...
$^{121}$Homer L. Dodge Department of Physics and Astronomy, University of Oklahoma, Norman OK; United States of America.\\
% ...
$^{138}$$^{(a)}$Departamento de F\'isica, Pontificia Universidad Cat\'olica de Chile, Santiago;$^{(b)}$Millennium Institute for Subatomic physics at high energy frontier (SAPHIR), Santiago;$^{(c)}$Instituto de Investigaci\'on Multidisciplinario en Ciencia y Tecnolog\'ia, y Departamento de F\'isica, Universidad de La Serena;$^{(d)}$Universidad Andres Bello, Department of Physics, Santiago;$^{(e)}$Instituto de Alta Investigaci\'on, Universidad de Tarapac\'a, Arica;$^{(f)}$Departamento de F\'isica, Universidad T\'ecnica Federico Santa Mar\'ia, Valpara\'iso; Chile.\\
% ...
$^{151}$Department of Physics, Technion, Israel Institute of Technology, Haifa; Israel.\\
$^{152}$Raymond and Beverly Sackler School of Physics and Astronomy, Tel Aviv University, Tel Aviv; Israel.\\
% ...
$^{a}$ Also Affiliated with an institute covered by a cooperation agreement with CERN.\\
$^{b}$ Also at An-Najah National University, Nablus; Palestine.\\
$^{c}$ Also at Borough of Manhattan Community College, City University of New York, New York NY; United States of America.\\
$^{d}$ Also at Center for High Energy Physics, Peking University; China.\\
$^{e}$ Also at Center for Interdisciplinary Research and Innovation (CIRI-AUTH), Thessaloniki; Greece.\\
$^{f}$ Also at Centro Studi e Ricerche Enrico Fermi; Italy.\\
% ...

\end{flushleft}


\end{document}